# Understanding high coercivity in ThMn$_{12}$-type Sm–Zr–Fe–Co–Ti permanent magnet powders through nanoscale analysis

Nikita Polin[1], Alexander M. Gabay[2], Chaoya Han[2], Christopher Chan[3], Se-Ho Kim[1,4], Chaoyang Ni[2], Oliver Gutfleisch[5], George C. Hadjipanayis[2], Baptiste Gault[1,6]

[1] Max Planck Institute for Sustainable Materials, Düsseldorf 40237, Germany

[2] University of Delaware, Newark, DE 19716, USA

[3] The Chemours Company, Chemours Discovery Hub, 201 Discovery Boulevard, Newark, DE 19713, US

[4] Department of Materials Science and Engineering, Korea University, Seoul, 02841, Republic of Korea

[5] Institute of Materials Science, Technische Universität Darmstadt, 64287 Darmstadt, Germany

[6] Department of Materials, Royal School of Mines, Imperial College, Prince Consort Road, London SW7 2BP, United Kingdom

# Graphical Abstract

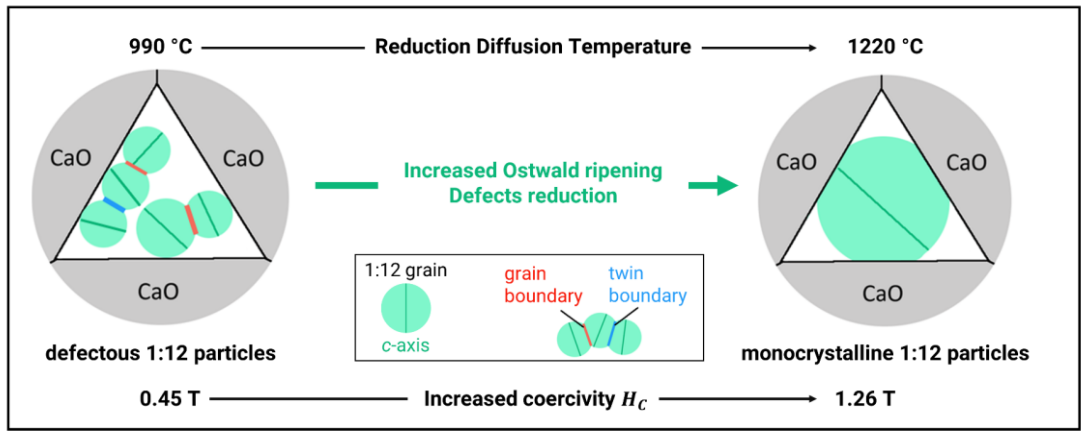

# Abstract

ThMn$_{12}$-type (Sm,Zr)$_1$(Fe,Co,Ti)$_{12}$ compounds show great potential for permanent magnets. Magnetically hard anisotropic powders prepared via reduction-diffusion exhibit a significant increase in coercivity from 0.45 T to 1.26 T as the processing temperature is raised from 990°C to 1220°C. Structural and microchemical analyses at high-resolution reveal that high-temperature processing annihilates grain boundaries (GBs) and reduces the density of twin boundaries (TBs), which are defects acting as weak links limiting the coercivity in the 1:12 system. Ostwald ripening is proposed as the mechanism behind the reduction of GB and TB densities at higher temperature, driven by the reduction in interfacial energy and enhancing atomic structural uniformity.

# Manuscript

Tetragonal ThMn$_{12}$-type compounds, commonly referred to as the 1:12 phase, with a space group *I*4/*mmm* and a composition of Sm$_{1-x}$Zr$_x$(Fe,Co)$_{12-y}$Ti$_y$, exhibit excellent intrinsic magnetic properties (high Curie temperature, high saturation magnetization, strong uniaxial magnetocrystalline anisotropy) and the lowest rare earth content among 3d-4f compounds, therefore qualifying as promising candidates to replace and potentially outperform widely used Nd-Fe-B-based magnets in medium- to high-temperature environments [1–5]. Partial Zr substitution for Sm is needed to increase the stability of the 1:12 compound, thus reducing the minimum necessary concentration of Ti and allowing for a larger saturation magnetization [1,6]. Achieving the superior performance requires, in addition to a high coercivity, a crystallographic alignment of the 1:12 crystallites. However, even relatively modest coercivities $\mu_0 H_c \approx 0.5$ T are usually obtained in the bulk 1:12 alloys based on Sm(Fe,Ti)$_{12}$ in the *nanocrystalline* state, which lacks the crystallographic texture [5,7–10] (we cite here only the most recent, 2022-2024, reports). Obtaining simultaneously the coercivity and the texture proved easier in the alloys based on the Sm(Fe,V)$_{12}$ compounds [11–13], but this compound has a much lower saturation magnetization. In 2021, Gabay *et al.* [14] reported a high $\mu_0 H_c$ = 1.26 T in (Sm,Zr)$_1$(Fe,Co,Ti)$_{12}$ particles prepared with a high-temperature calciothermic reduction. Being nearly monocrystalline, these particles can easily be textured with a magnetic field. Even though they are not immediately suitable for manufacturing the fully dense magnets (which usually also need to contain the specific minority phases), a better understanding of the particles' high coercivity is deemed to be important for the development of such magnets.

Poor coercivity of the 1:12 alloys has been attributed to the easy nucleation of a reversed magnetization at grain boundaries (GBs) [15] or at twin boundaries (TBs) within the grains of the 1:12 phase [16,17]. Gabay *et al.* demonstrated that increasing reduction-diffusion annealing temperature from 990°C to 1220°C in $Sm_{1-x}Zr_x(Fe,Co)_{12-y}Ti_y$ significantly enhanced coercivity $\mu_0 H_C$ from 0.45 to 1.26T [14]. This improvement was linked to the higher monocrystalline quality of the grains, with better separated grains (fewer GBs) and, as hypothesized by the authors, a lower density of structural defects [14].

Here, we build on these previous efforts to investigate the structure and microchemistry of the same two samples annealed at 990°C (low coercivity) and 1220 °C (high coercivity) using transmission electron microscopy (TEM) and atom probe tomography (APT) [14]. Our goal is to address the following research questions when comparing these two samples: (1) How do their microstructures differ, particularly in regard to observed defects? (2) Can we explain the microstructure differences based on synthesis conditions? (3) Do these microstructure differences explain the large difference in coercivity? The 990 °C sample contains numerous GBs and TBs that were found to be eliminated or significantly reduced in the 1220°C sample. The improved monocrystalline quality of the 1220°C sample results in coercivity enhancement, identifying these GBs and TBs as critical weak links for the demagnetization of the 1:12 Sm–Zr–Fe–Co–Ti system. We propose a model in which the disappearance of GBs and most TBs at higher temperatures is explained by competitive particle growth, known as Ostwald ripening, in confined spaces between the grains of the CaO dispersant [18].

Magnetically hard, anisotropic powders were prepared by subjecting a mixture of elemental oxides ($Sm_2O_3$, $ZrO_2$, $Fe_2O_3$, $TiO_2$), Co, Ca (reducing agent) and CaO (dispersant) to a sequence of high-energy ball milling, reduction-diffusion at 990–1220 °C, and repeated washing, as described in detail in Ref. [14]. Typical BSE SEM micrographs of the as-annealed (i.e., not washed) mixture are shown in the Supplement (**Figure S1**). The key results from magnetic (VSM) and structural (XRD, SEM) characterization reported in Ref. [14] are summarized in **Table 1**.

Samples used for TEM observations were prepared using focused ion beam (FIB) to mill particles embedded in resin. This preparation and initial SEM imaging were performed with a Zeiss Auriga 60 CrossBeam. Bright-field (BF) TEM images and high-angle annular dark-field (HAADF) scanning TEM (STEM) images were acquired using both a Thermofisher Scientific Talos F200C and Themis G3 S/TEM instrument with $C_s$ probe-correction. Compositional mapping was performed using a 450 pm STEM probe with the SuperX EDS system on the Themis. The mapping data consisted of images with a resolution of 512 x 512 pixels, with collection times ranging from two to eight hours per dataset.

For nanoscale composition analysis, the specimen preparation workflow is outlined in the supplement (**Figure S2**). The powder particles were embedded in a Ni film by co-electrodeposition [19,20] performed for 600 s at a constant current of 19 mA on a Cu stub. Embedded particles containing regions were identified as protruding islands on the Ni film surface, from which needle-shaped specimens were prepared by a Ga-ion focused ion beam (FIB) (Dual-Beam Helios Nanolab 600i, FEI). The preparation followed the procedure outlined by Thompson et al. [21], with final milling conducted using a low energy (5 keV) Ga beam to remove beam-damaged regions. These specimens were analyzed using a CAMECA LEAP 5000 XR reflectron-equipped atom probe, operating at 40 K under ultra-high vacuum ($10^{-10}$ mbar), with pulsed UV laser (355 nm wavelength, 10 ps pulse duration, 45 − 80 pJ pulse energy, 125 kHz pulse rate) and a detection rate of 1–2 ions per 100 pulses on average. Datasets containing at least 15 million ions were analyzed by using CAMECA AP Suite v.6.3.

Table 1 Samples investigated in this publication, taken from Gabay *et al*. [14]. Distributions of particle size and grain sizes can be found in the supplement of [14].

| Sample | Sample 1 | Sample 2 |
|---|---|---|
| Reduction-diffusion annealing temperature $T_a$ (°C) | 990 | 1220 |
| Coercivity $\mu_0 H_C$ (T) | 0.45 | 1.26 |
| Remanent magnetization $\sigma_r$ (A·m$^2$/kg) | 84 | 124 |
| Particle size: particle length $L$; particle width $W$, mean values (µm) | 1.06; 0.50 | 0.65, 0.53 |
| Grain size, volume weighted $(<D^3>)^{1/3}$ (µm) | 0.35 | 0.54 |
| Phases by XRD | ThMn$_{12}$ type ("1:12", balance), W type ("bcc", 3.1 vol. %) | ThMn$_{12}$ type ("1:12", balance) W type ("bcc", 4.7 vol. %.) NaCl type ("CaO", 9.2 vol. %.) CaF$_2$-type [(Zr,Ti)H$_{1.8}$, 3.0 vol. %] |

**Figure 1a** shows a low magnification HAADF STEM image of the powder heated at 990 °C. The particles exhibit a broad aspect ratio distribution (also cf. Supplement of Ref. [14]) and generally contain multiple grains. Approximately 13% of the particles contain twins, which are identified by sharp interfaces with abrupt contrast changes. Representative examples are highlighted with yellow arrows in the HAADF image (**Figure 1a**). Additional particles are displayed in the supplementary (**Figure S3a-b**). A single particle containing several grains, numbered in the medium magnification HAADF STEM image shown in **Figure 1b**, is selected for further characterization. The region of interest includes a GB triple junction, with GBs marked by white arrows in the HAADF STEM image.

High magnification HAADF STEM (**Figure 1c**) and STEM EDX (**Figure 1c1-c5**) imaging of the GB junction reveals similar compositions in grains #1 and #2, while grain #3 is enriched with Ti and Zr, with concentrations deviating by up to 10 at.% from the other grains (see EDX line profiles in supplementary, **Figure S3c**); The GBs between the grains #1-#4 and #2-#4 are partially decorated by Zr and Ti on their exterior surface. Grains #1 and #2 are identified as the 1:12 phase, while grain #3 belongs to the hexagonal C14 Laves phase [22] with the formula $(Fe,Co)_{2+\delta}(Ti,Zr)_{1-\delta}$ and lattice parameters $a$ = 0.4757 nm and $c$ = 0.7829 nm, as determined by the bright field TEM image (**Figure 1d**), FFT patterns from respective grains (**Figure 1d1-d3**) and EDX analysis (supplementary **Figure S3c**). Note that the C14 Laves phase was not detected by XRD by Gabay *et al.* [14], indicating a low volume fraction of this higher local Ti+Zr concentration.

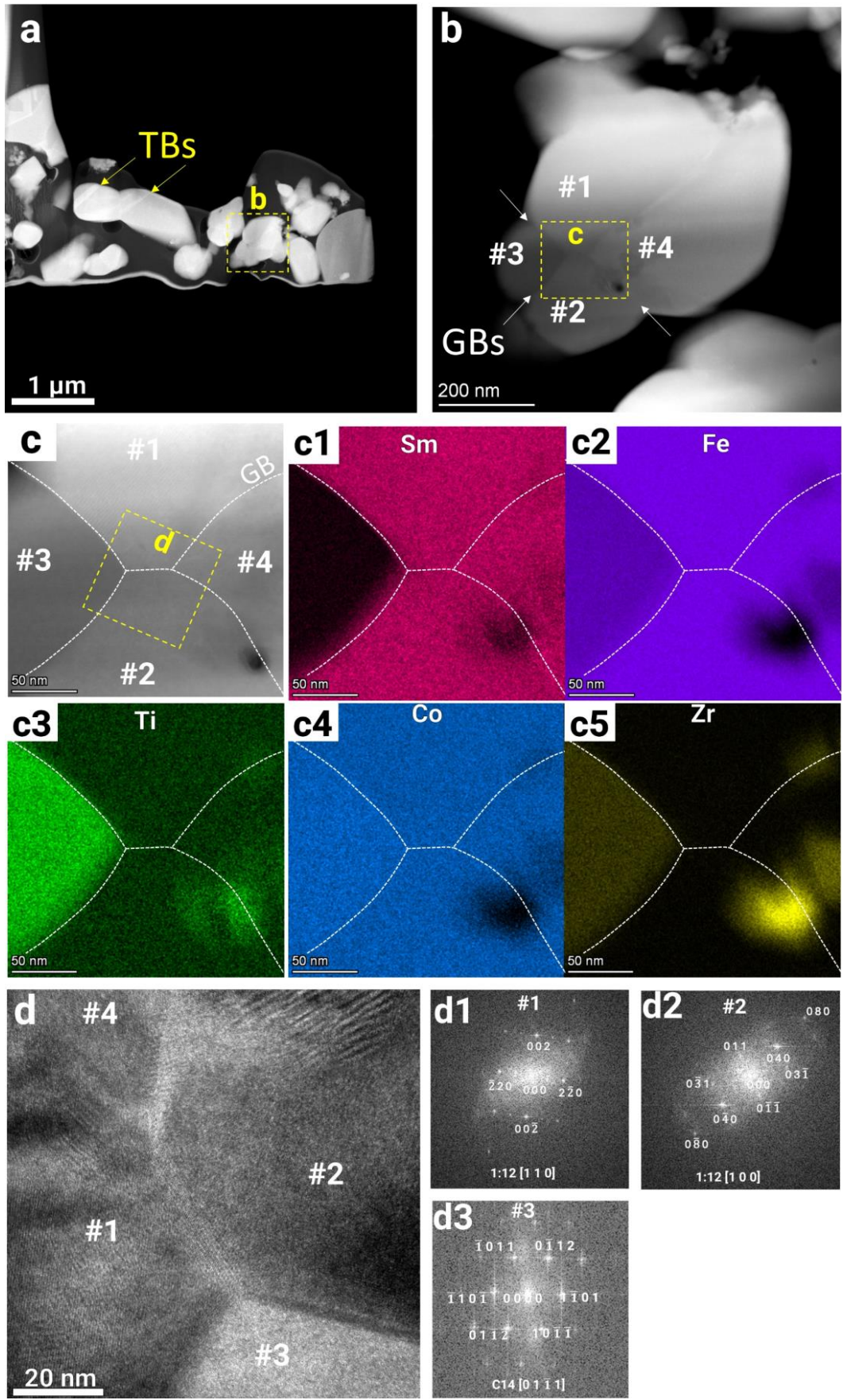

**Figure 1** TEM investigation of Sm–Zr–Fe–Co–Ti particles synthesized at 990 °C: **(a-c)** HAADF STEM at different magnifications showing assembly of particles, a particle of interest and its GB triple junction region, respectively. **(c1-c5)** STEM EDX images from region shown in **(c)**. **(d)** Bright-field (BF) TEM and **(d1-d3)** FFT patterns extracted from respective grains in **(d)**. Yellow and white arrows highlight TBs and GBs, respectively. In images (c,c1-c5) the GBs are marked by white dotted lines.

**Figure 2a** displays a low-magnification BF TEM image of the powder heated at 1220 °C. The particles exhibit a narrow aspect ratio distribution (also cf. Supplement of Ref. [14]), characterized by distinctive facets, in line with SEM results from Ref. [14]. These particles are composed of individual grains and are mostly free of GBs, except for around 8% of them (10 out of 140 particles), which display residual twins, as seen in this figure and other TEM images (cf. supplementary **Figure S3d**). A particle containing a twin boundary is imaged at medium and high magnification by bright field TEM, **Figure 2b-c**, and it is identified as 1:12 phase by FFT in **Figure 2d**.

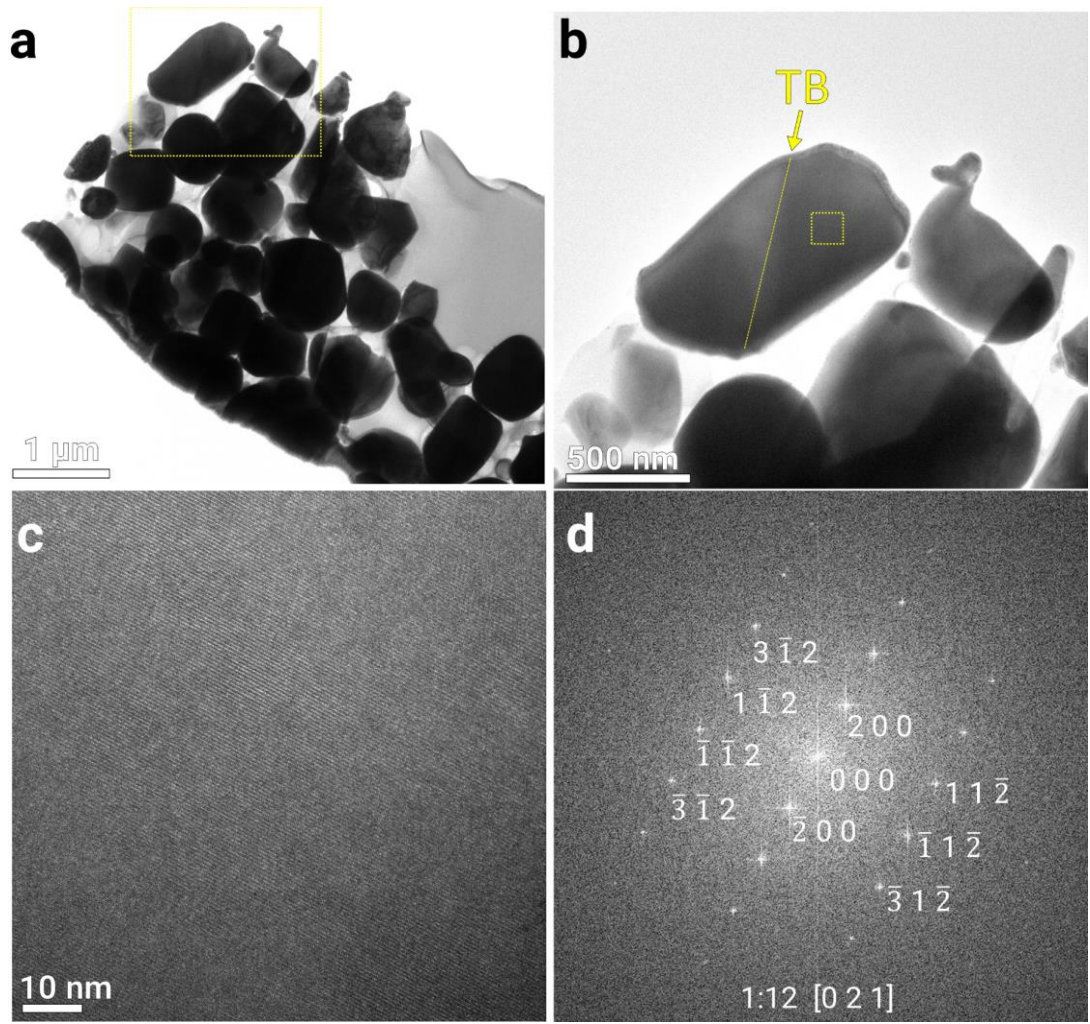

**Figure 2** TEM investigation of Sm–Zr–Fe–Co–Ti particles synthesized at 1220 °C. **(a-c)** Bright field TEM images at increasing magnifications. **(d)** FFT from image **(c)** in [021] zone axis of the 1:12 phase. The TB is marked by a yellow dashed line in image (b).

The 3D reconstruction resulting from an APT analysis of a representative 990 °C particle is displayed in **Figure 3a**. Two grains can be distinguished, separated by a Zr-rich 2D feature, as further visualized by a contour map in **Figure 3b**. Due to the compositional difference, this feature can be assigned to a GB, separating two grains, consistent with TEM observations. The composition of $Sm_{7.2}Zr_{0.9}Fe_{69.8}Ti_{6.1}Co_{16.1}$ found in grain #1 is representative of the 990 °C sample, roughly matching 1:12 stoichiometry (**Table 2**). For grain #2, the 1D profile shown in **Figure 3c** reveals a reduction in Zr (-0.3 at.%) and Ti (-0.3 at.%) and an increase in Sm (+0.3 at.%), Co (+0.2 at.%) and Fe (+0.1 at.%) compared to grain #1, also approximately aligning with a 1:12 stoichiometry. The ≈2 nm thin GB is enriched in Co (+1 at.%) and Zr (+0.5 at.%) and depleted in Fe (-0.9 at.%) and Ti (-0.6 at.%) compared to grain #1.

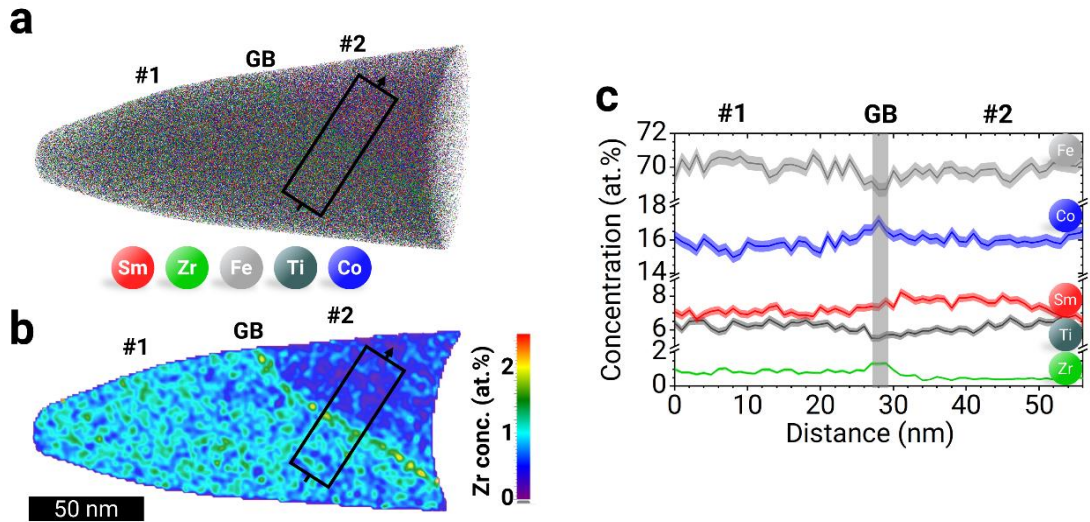

**Figure 3** APT investigation of Sm–Zr–Fe–Co–Ti particles synthesized at 990 °C. **a** 3D reconstruction of APT data. **b** 2D Zr composition profile. **c** 1D profile calculated from a cylinder (60nm long, 20nm in diameter) shown in **a**. Note the interruptions of the y-axis of the 1D profile.

**Table 2** Averaged compositions $c$ and standard deviations from several APT measurements for Sm–Zr–Fe–Co–Ti alloys synthesized at different temperatures. Note that for the stochiometric compound $(Sm,Zr)_1(Fe,Co,Ti)_{12}$, the expected concentrations are 7.7 at.% for Sm+Zr and 92.3 at.% for Fe+Co+Ti

| Atom | $C_{990°C}$ | $C_{1220°C}$ | $C_{1220°C} - C_{990°C}$ |
|---|---|---|---|
| Fe | 69.8±0.1% | 71.6±0.3% | 1.8±0.3% |
| Co | 16.1±0.1% | 15.5±0.1% | -0.6±0.1% |
| Sm | 7.2±0.1% | 5.6±0.2% | -1.6±0.2% |
| Ti | 6.1±0.1% | 6.2±0.1% | 0.1±0.1% |
| Zr | 0.9±0.1% | 1.1±0.1% | 0.2±0.1% |
| Sm+Zr | 8.1±0.1% | 6.7±0.2% | -1.4±0.2% |
| Fe+Co+Ti | 91.9±0.1% | 93.3±0.3% | 1.4±0.3% |

**Figure 4a** is a representative 3D atom map from the APT analysis of the particle heated at 1220 °C. Ni was used as the embedding material for the powder in the electrodeposition process [23,24], and it is found at the start of the APT analysis, indicating that we have captured the particle's interface with metallic matrix. This is consistent with particles treated at 990 °C (see supplementary **Figure S4a**). As shown by the 1D profile in **Figure 4b**, the composition gradually changes from pure Ni to the uniform composition of the particle. The measured particle composition of $Sm_{5.4}Zr_{1.2}Fe_{71.8}Ti_{6.1}Co_{15.5}$ for the 1220 °C sample (**Table 2**) slightly deviates from 1:12 stoichiometry. The composition uniformity and absence of defects, especially of GBs, further support the assumption of mostly monocrystalline 1220 °C particles, corroborating TEM results.

Less commonly, in one out of four APT-probed particles, multiple phases - including 1:12, $Ti(Fe,Co)_2$ and $\alpha(Fe,Co)$ - were detected, which might have formed due to local fluctuations of Sm, Ti and Zr. The multiphase particle was clearly separated by Ni from the main phase particle, as shown in the supplement (**Figure S4b**). The presence of such multiphase particles might explain the non-uniformity of the demagnetization curve (supplement **Figure S5**).

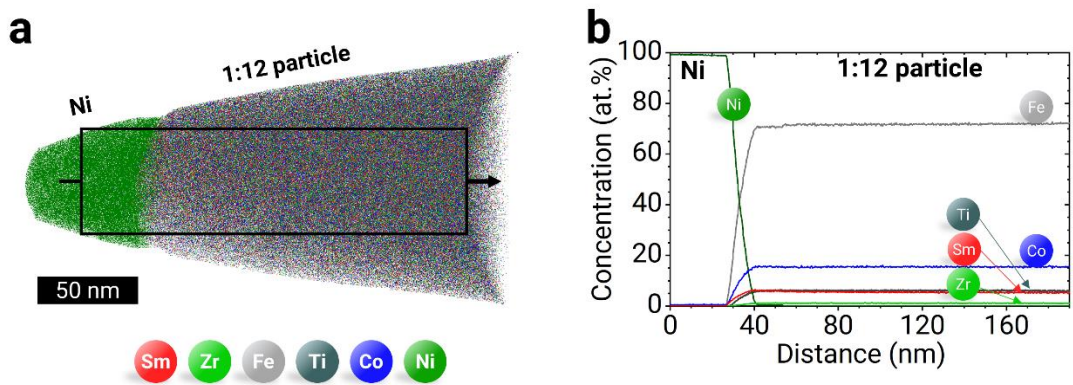

**Figure 4** APT investigation of Sm–Zr–Fe–Co–Ti particles synthesized at 1220 °C. **a** 3D reconstruction of APT data showing embedding material Ni and the 1:12 particle. **b** 1D profile calculated from a cylinder (190nm long, 60nm in diameter) indicated in **a**.

Drawing from our TEM and APT observations, we now address the initially raised questions regarding Sm–Zr–Fe–Co–Ti particles synthesized at 990 °C and at 1220 °C: (1) How do their microstructures differ, particularly in regard to observed defects? (2) Can we explain the microstructure differences based on synthesis conditions? (3) Do these microstructure differences explain the large difference in coercivity? For answering these questions, we propose a model for 1:12 Sm–Zr–Fe–Co–Ti particle growth during annealing which occurs in a void between grains of the CaO dispersant, as schematically sketched in

**Figure 5.** Typical BSE SEM micrographs of such as-annealed 1:12 particles surrounded by CaO crystals are shown in the Supplement (**Figure S1**).

(1) *Microstructure*: Apart from slight composition differences, the main distinction is a higher monocrystallinity of the main phase in the particles synthesized at 1220 °C compared to those synthesized at 990 °C. This is characterized by significantly fewer twin boundaries (≈60% less) and the absence of GBs. In both samples, more minority phases were observed than detected with the XRD by Gabay *et al*. [14]. The distinctive facets of the particles were only observed in particles synthesized at 1220 °C, as observed by SEM in Ref. [14] and TEM in this study.

(2) *Synthesis conditions*: The observed differences in microstructures can be explained by the mechanism of Ostwald ripening, where larger particles grow at the expense of smaller ones, thereby reducing the system's overall interfacial energy [18,25]. Ostwald ripening is typically more pronounced at higher temperatures for metallic systems, resulting, for a fixed annealing time, in a smaller number of particles with larger average size [26–28].

During reduction-diffusion annealing, Sm–Zr–Fe–Co–Ti particles nucleate heterogeneously at CaO grain surfaces and subsequently grow, as illustrated in **Figure 5**. At a lower temperature of 990 °C, the reduced atomic diffusion rates limit Ostwald ripening, resulting in a regime dominated by nucleation and growth. As the single-crystal particles grow, they come into contact and merge, forming GBs or, in the case of specific particle orientations, TBs. In addition to the formation by the merging of two particles, nucleation and structural fluctuation during heat treatment can generate TBs in single particles, as it is commonly observed in metallic nanoparticles [29].

In contrast, at a higher temperature of 1220 °C, Ostwald ripening is more pronounced due to an increased atomic mobility. In this case, after nucleation, only the largest particles continue growing, while smaller particles shrink, hindering the formation of GBs and TBs via the particle merging. As a result, mostly monocrystalline particles in voids of CaO matrix form. Twinned particles (around 8 %) form in rare cases, where a single particle acquires a TB by structural fluctuation and grows more rapidly than competing particles. These TBs are not "healed" during annealing most likely due to the short treatment time of 5 min, preventing the system from reaching thermodynamic equilibrium.

It is important to note that Ostwald ripening, as described here, occurs within confined spaces, which qualitatively differs from free-space Ostwald ripening due to mechanical interactions between the particles and the support [30]. In our case, the particles coarsen within the CaO matrix and, by coming into mechanical contact with CaO interfaces, develop distinctive facets. This finding further demonstrates a more pronounced Ostwald ripening at higher temperatures.

For the 990°C sample, only a single GB could be captured in three APT measurements. Targeted specimen preparation at a GB remains challenging considering the size of the particles and the volume measured by APT. Note that the APT data did not show sufficient crystallographic information to allow further analysis of the grains orientation and nature of the GB [31]. Since the GBs form by attachment of randomly oriented particles with a non well-defined shape, they are expected to primarily be random high-angle GBs [32]. Twins can either be related to a few rare cases where facetted particles attached along specific sets of planes, or formed from moving defects during the heat treatment. The composition of this GB can hence be seen as representative of a random high-angle GB.

(3) *Coercivity*: The initial magnetization curves in [14] show that 1:12 phase Sm–Zr–Fe–Co–Ti particles are nucleation-type magnets [33]. The increase in coercivity is thus likely associated with the removal of the GBs and a lower number density of TBs, as both defects are known to be weak links, where domains with reversed magnetization can more easily nucleate. The detrimental effects of TBs [16] and GBs [15] on coercivity have been clearly demonstrated for bulk $Sm(Fe,Ti)_{12}$. Our study suggests that the same is true for Sm–Zr–Fe–Co–Ti anisotropic powders. Even though the GB specific composition probed by APT belongs to the 1:12 homogeneity range [34], it may have a different crystal structure (e.g. TbCu$_7$-type [10]) or even be amorphous, and as a consequence exhibit a lower anisotropy constant and act as a weaker link. Hence, further studies including TEM, magnetic domain imaging, ab initio and micromagnetic simulations, are needed to further clarify GB structure, intrinsic properties and behavior during demagnetization for anisotropic powders. The simultaneous increase of grain size and coercivity (see Table 1) is contrary to the usually observed trend [35,36], indicating that, in our case, the detrimental impact of defects on coercivity outweighs any potential benefits from smaller grain size effect [14].

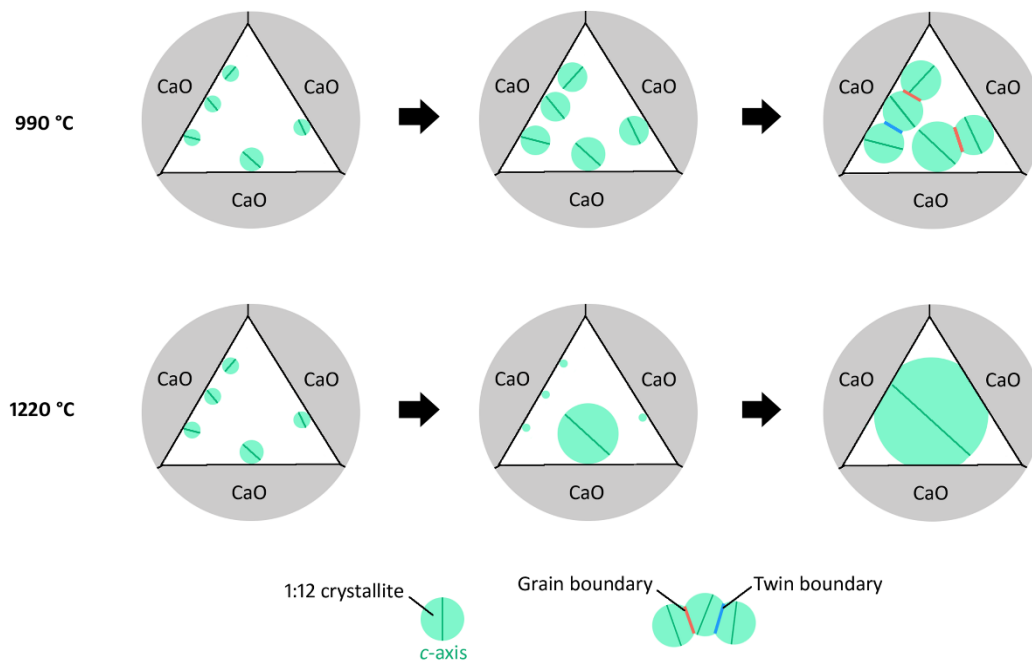

**Figure 5** Concept sketch explaining the different microstructures of 1:12 Sm–Zr–Fe–Co–Ti particles (green) synthesized at reduction diffusion temperatures of 990 °C and 1220 °C in a void between CaO grains (support matrix, grey). GBs and TBs are marked in red and blue, respectively, *c*-axes indicated by dark green lines.

In conclusion, our combination of high resolution TEM and APT techniques suggests that grain boundaries and twin boundaries act as magnetic weak links in nucleation-type $Sm_{1-x}Zr_x(Fe,Co)_{12-y}Ti_y$ permanent magnet powders synthesized by reduction-diffusion, limiting their coercivity. Increasing the reduction-diffusion annealing temperature likely promotes Ostwald ripening of the 1:12 particles, which removes the grain boundaries and reduces the incidence of twin boundaries. The magnetically anisotropic monocrystalline powders produced by this method show the promise as precursors for the development of bonded and sintered 1:12 permanent magnets.

## Acknowledgements

We thank Uwe Tezins, Christian Broß, and Andreas Sturm for their support to the FIB & APT facilities at MPIE. NP is grateful for the financial support from International Max Planck Research School for Sustainable Metallurgy (IMPRS-SusMet). C.H. and C.N. acknowledge partial support by the U.S. Department of Energy, Office of Science, Office of Basic Energy Sciences under Award Number DE SC0022168. S.-H.K acknowledge supports from KIAT-MOTIE (P0023676, HRD Program for Industrial Innovation and RS-2024-00431836, Technology Innovation Program). This work was also supported by the Deutsche Forschungsgemeinschaft (DFG, German Research Foundation), Project ID No. 405553726, CRC/TRR 270.

## Declaration of Competing Interest

The authors declare that they have no known competing financial interests or personal relationships that could have appeared to influence the work reported in this paper.

# Supplementary Material:

# Understanding High Coercivity in ThMn$_{12}$-type-Type Sm–Zr–Fe–Co–Ti Permanent Magnet Powders through Nanoscale Analysis

Nikita Polin[1], Alexander M. Gabay[2], Chaoya Han[2], Christopher Chan[3], Se-Ho Kim[1,4], Chaoyang Ni[2], Oliver Gutfleisch[5], George C. Hadjipanayis[2], Baptiste Gault[1,6]

[1] Max Planck Institute for Sustainable Materials, Düsseldorf 40237, Germany

[2] University of Delaware, Newark, DE 19716, USA

[3] The Chemours Company, Chemours Discovery Hub, 201 Discovery Boulevard, Newark, DE 19713, US

[4] Department of Materials Science and Engineering, Korea University, Seoul, 02841, Republic of Korea

[5] Institute of Materials Science, Technische Universität Darmstadt, 64287 Darmstadt, Germany

[6] Department of Materials, Royal School of Mines, Imperial College, Prince Consort Road, London SW7 2BP, United Kingdom

# A. Material synthesis and processing

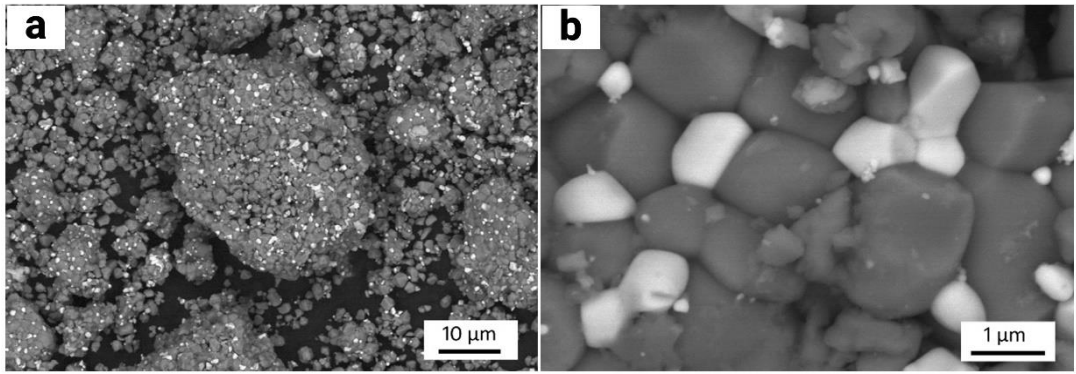

**Figure S1** Typical BSE SEM micrographs of as-annealed (i.e., not washed) composite with bright 1:12 particles surrounded by darker CaO crystals at (**a**) low and (**b**) high magnification. Brighter contrast of the 1:12 particles is due to a larger average atomic number of the constituent elements. The reduction annealing temperature was 1200°C.

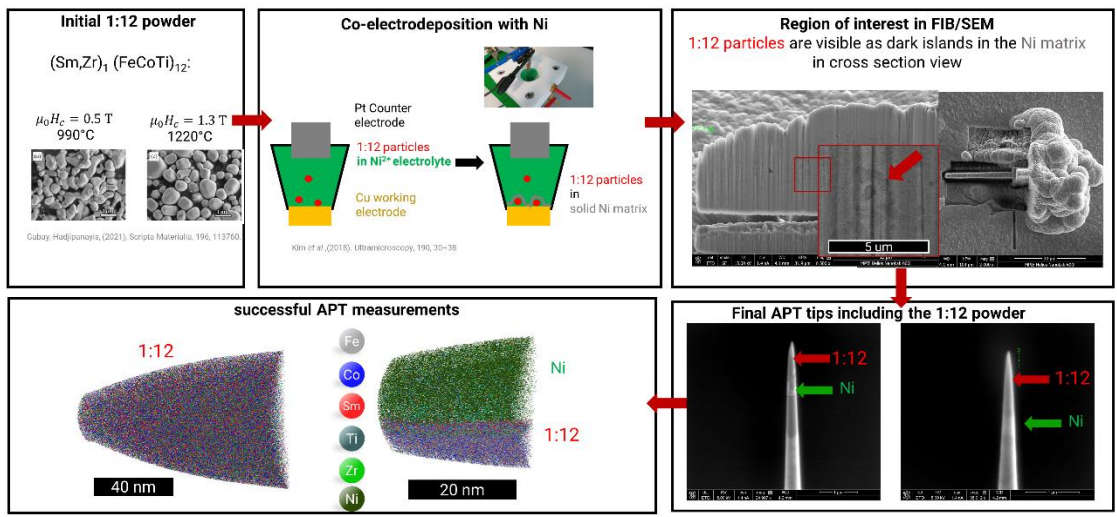

**Figure S2** Workflow of APT tips preparation using the electrodeposition method.

## B. Microstructure analysis

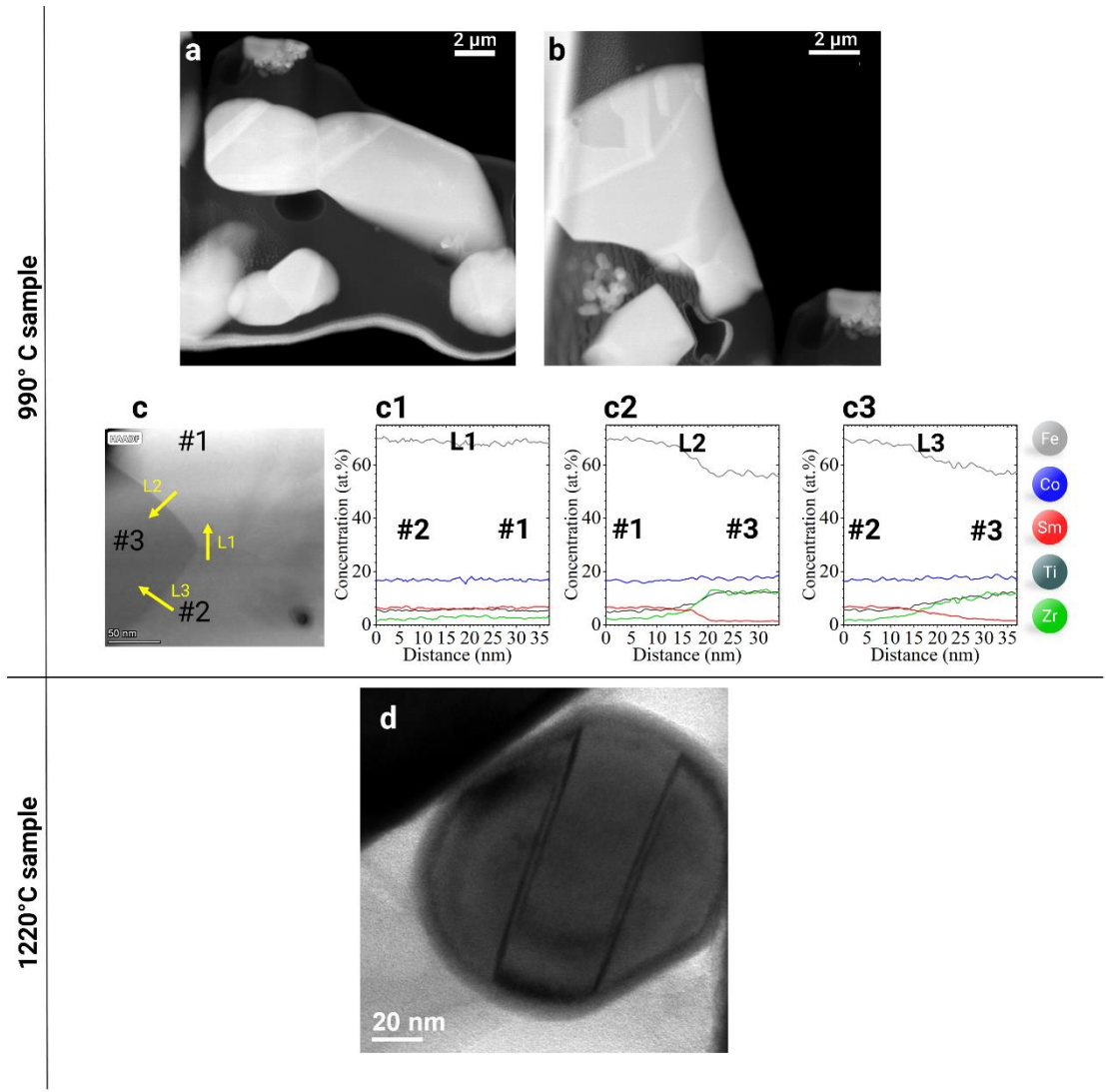

**Figure S3** TEM investigation of 990 °C **(a-c)** and 1220 °C sample **(d)**. **(a-b)** HAADF STEM images of particles containing twin boundaries. **(c)** HAADF STEM of a GB junction and **(c1-c3)** corresponding EDX profiles along 3 GBs enumerated L1-L3, whereas grain #1 and #2 are 1:12 phase, #3 is C14 Laves phase. **(d)** BF TEM image of a particle containing two twin boundaries.

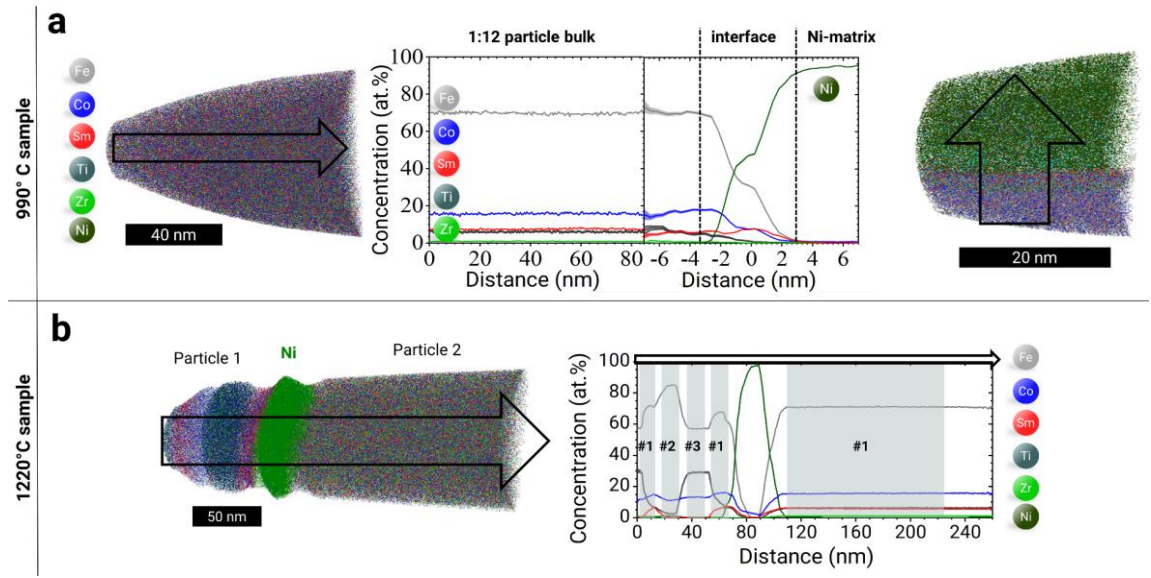

**Figure S4** APT datasets of the of **(a)** 990°C and **(b)** 1220°C particles with concentration profiles. **(a)** (left) 1:12 particle without a Ni-interface and corresponding a 1D profile; (right) 1:12 particle with a Ni-interface and corresponding proxigram profile. After the transition region, across the 7nm wide interface, the bulk composition of the particle is restored, demonstrating that Ni-electrodeposition does not alter the bulk composition of the Sm-Ti-Fe-Co-Zr particles. Only at the surface of the particles, some elements likely diffused into the interface region during the electrodeposition process. **(b)** APT dataset of the 1220°C sample, 2 particles separated by embedding material Ni; particle 1 contains the phases 1:12 (#1), $\alpha$(Fe,Co) (#2) and Ti(Fe,Co)$_2$ (#3), and particle 2 contains 1:12 phase.

## C. Magnetic properties

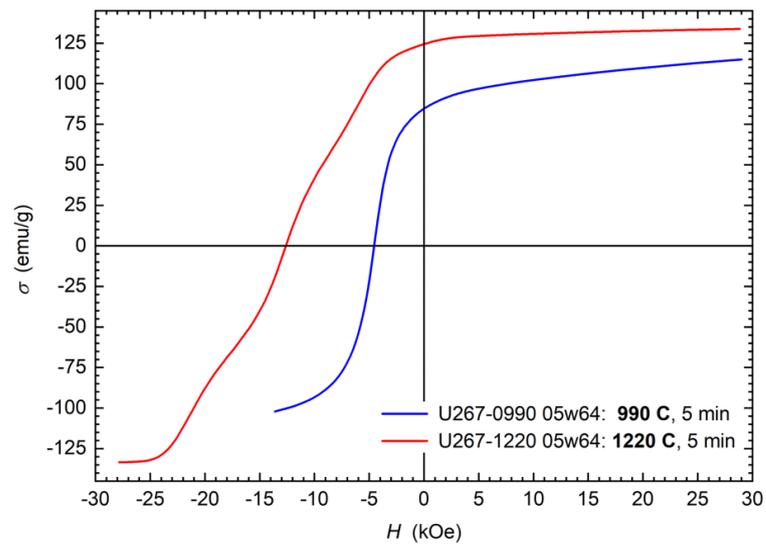

**Figure S5** Demagnetization curves of investigated samples. Note, that for the 1220 °C sample, the curve is non-uniform, indicating the sample contains particles od slightly different coercivities